# To mechanics of deformation, flow, and fracture

S.L. Arsenjev[1]

Physical-Technical Group
Dobroljubova Street, 2, 29, Pavlograd Town, 51400 Ukraine

It is stated in the main in essence new approach to mechanics of the stressed state of the solid body from statistically isotropic material and the homogeneous liquid dynamics. The approach essence is in the detected property of the core-shell spontaneous structurization of internal energy of the solid and liquid bodies in its natural state and under action of external forces. The method elements of construction of physically adequate model of the stressed state of the solid and liquid bodies, reproduced exactly its behavior on the stages of elastic and plastic deformation, flow and fracture, are stated. It is adduced a number of the examples of the stressed state construction of the simple form bodies under action of its tension, compression, torsion and at its contact interaction. For the first time it is adduced structure of the principal – normal - stresses in cylindrical bar under action of the torsion moment. The detected property and the developed method is one of necessary bases for construction of physically adequate mathematical model of the stressed state of the body and fluid in contrast to traditional approach.

**Introduction**
The mechanics analysis of a forming of the neck in cylindrical specimen from statistically isotropic and elastic-plastic material, in particular, from the mild steel under action of the longitudinal (axial) tension forces, stated in author's article [1], had led to necessity to acceptance of two following suppositions: firstly, on distribution of trajectories of principal – normal – stresses in this specimen in the kind of cylindrical helices with its angle $45^0$ and, secondly, on a self-sructurization of the specimen volume into its core and shell parts. These two suppositions had allowed to account rationally for all features of the stress and strain states of such specimen at all stages of its testing, including its fracture. According to the second from the accepted suppositions, the average-statistical quantity of diameter of the specimen core part makes up $\sqrt{2}/2 = 0.707...$ of its external diameter. This ratio remains to be correct for the neck diameter in comparison with the specimen external diameter, and the same ratio is correct for the neck core part in comparison with the neck diameter. Deviation from the indicated quantity not exceeds 20 % for the different sorts of structural steel [2]. The adduced ratio in combination with the first from the accepted suppositions has allowed geometrically strict and physically adequate to construct the specimen form at all stages of its loading right up to its fracture with use only two geometrical surfaces: the hollowed orthogonal – right – hyperboloid and evolventoid of rotation. Such approach has allowed to ascertain that the necking process is a phenomenon of the local straightening of helical trajectories of the principal stresses and, accordingly to it, of a transformation of rectilinear generatrix on the specimen cylindrical surface into a hyperbola in the very narrow zone of the specimen neck. Presentation of the cylindrical volume in the kind of a structure, contained in itself the core and shell parts, has allowed to discover the real three-dimensional structure of the stressed state in the cylindrical specimen under simple tension and in that way to show that the one-dimensional stressed state as only stress in three-dimensional body is an abstraction, devoid of physical sense. Physically substantial and mathematically strict solution of a problem on simple tension of cylindrical specimen and some consequences, stated in previous article [1], can be, in author's opinion, applied for physically adequate solution of problems on the stressed-strained state of the specimens with different form of its cross-section under not only the tension test, but also under axial compression, torsion and some other kind of a testing under both static and dynamic load.

---
[1] Phone: (+38 05632) 40596 (home Rus.)
E-mail: ptglecus@yahoo.com

The principal problem of a generalized approach to a problem solution on the stressed-strained state of the specimens under the numbered kinds of a testing consists in possibility to elucidate nature and features of the spontaneous structurization of statistically isotropic material in a specimen volume before application to it a test load and the influence of this phenomenon onto the stress distribution in the specimen under action of external forces.

**Approach**

The approach essence to the stated problem solution consists in the following theses:
- a solid body in its initial – unloaded – state keeps its form and its volume owing to its internal energy, ensuring a stationary connection of micro particles, constituting the solid body material; an intensity of this connection is characterized by the linear modulus and the volumetric modulus of elasticity in dimensionality $kgf/m^2 \equiv kgf \cdot m/m^3$ as a volumetric density of the internal connection energy of the body material; thus, the solid body is in the own stationary stressed state, keeping its initial form and volume in a lack of external influences;
- the stationary connection of the body material particles has the essentially oscillatory character and thereby stipulates a pulsatory – oscillatory – character of the own initial stressed state of the body from the simplest – spherical – form with its radial and shear pulsations to a combination of radial, longitudinal, torsion, flexural pulsations in the cylindrical form body and up to the more complicated combinations of the pulsations in the complicated form bodies; the stress waves, moving to the body surface and refracting from it, form the pulsatory system of initial stationary stresses, corresponding structurally to the body form; in the result of it the initial stressed state of the body is essentially three-dimensional accordingly to the body form and this state has a cyclical character with the oscillation frequency, determined by a sonic wave velocity in the body material, by the oscillation mode and by the body characteristic dimensions;
- any mechanical or thermal influence from without onto the essentially three-dimensional body can be able only quantitatively to change a correlation of components of its own initial three-dimensional stressed state.

In the given article as a solid body is considered the elastic-plastic body from statistically isotropic material, as far as such body under influence of external forces demonstrates ability to:
- elastic (reversible) deformation;
- the vacancy-dislocation adaptation of its material microstructure to its stressed state on the plastic (irreversible) deformation stage;
- the local and as a whole flow of such body material;
- a loss of continuity on its fracture stage.

A body from brittle material, possessing by its tension strength in 2…5 times less than its compression strength, demonstrates only little part of the plastic body behavior when other things being equal. At the same time the well-known experimental researches [3], carried out by T. Karman (1911), had showed that the cylindrical specimens from marble demonstrate the plastic deformation under action of a combination of axial compression with lateral hydrostatic pressure.

**Solution**

As regards a distribution of trajectories of principal stresses in the kind of two families of cylindrical – left- and right-handed – helices, its existence is, although indirectly, corroborated by Lueders lines on the cylindrical specimen surface on a stage of its plastic deformation under action of the axial tension force.

A question on a possibility of the spontaneous structurization of the initial stressed state in isotropic three-dimensional body is, in the given article, considered on the examples of bodies of some simple forms.

Spherical body. One of the basic modes of its initial spontaneous oscillations is realized in the kind of radial oscillation. In this case an interaction of the own oscillation waves, moving to the sphere surface, with the refracted waves, moving to the sphere centre, results in that the near-walled layer of the sphere material will be oscillated in opposite phase to the sphere central part. In that way the sphere volume contains its core and shell parts. Such oscillatory system can be presented in the kind of two bodies connected by spring. It is known [4] that, firstly, the free oscillations of such mechanical system are going on without displacement of the system mass centre and, secondly, both bodies move in opposite phases with the same frequency. When mass of these bodies is unequal then its oscillation amplitudes is inversely proportional to its mass. In particular case, when mass of these bodies is the same, its amplitudes will be also equaled. As applied to the spherical body, this particular case corresponds to the continuity condition of its material in a contact boundary of this body core and shell parts. For the sphere isotropic material, radius, dividing it on its core and shell parts, is determined by equality of volumes of these parts

$$V_c = \frac{4}{3}\pi \cdot R_c^3; \qquad V_s = 2V_c = \frac{4}{3}\pi \cdot R_s^3,$$

whence

$$R_c/R_s = 1/\sqrt[3]{2} = 0.7937...,$$

where $R_c$ is radius of the sphere core part, dividing the sphere volume into two equal parts and $R_s$ is radius of the sphere. Accordingly, a thickness of the sphere shell part makes up approximately 20.6 % from the sphere radius ($R_s$).

In a process of these stationary radial oscillations, the stress in the sphere core is changed periodically – during one half-period – from the volumetric compression to the volumetric tension. Accordingly to it the stresses in the sphere shell – the hoop stresses in two perpendicular directions and radial stress – are also changed periodically from a tension to a compression. In that way, it should be supposed the continual spherical body as an aggregate, consisting from two elastically connected parts, when an analysis of the stressed-strained state of such body under action of external load is carried out.

In a case of rotational oscillations, the sphere core radius is determined by an equality of the inertia moments of its core and shell parts.

Cylindrical body of a finite length. One of the basic modes of its initial spontaneous oscillations of such body is also realized in the kind of radial oscillations. Acting analogically to the previous case and supposing an equality of mass (volume) of its core and shell parts, one can determine radius of the cylinder core part

$$V_c = \pi \cdot R_c^2 \cdot L; \qquad V_{cb} = 2V_c = \pi \cdot R_{cb}^2 \cdot L,$$

whence

$$R_c/R_{cb} = 1/\sqrt{2} = 0.707...,$$

where $R_c$ is the cylinder radius, dividing its volume into two equal parts, i.e. radius of the cylinder core part; $R_{cb}$ and $L$ are radius and length of the cylinder respectively.

In this mode of initial own oscillations, the cylinder shell is periodically compressing its core. In the result, the core is lengthened and one is tensioning the shell in longitudinal direction in one half-period of the oscillations. The next one half-period of the own oscillations, the cylinder core is shorten and, dilating simultaneously in radial direction, one is arousing the tension hoop stress and the compression longitudinal stress in the cylinder shell. These own oscillations of the cylinder core and shell parts correspond to the oscillatory nature of the longitudinal and transversal connections of micro particles, constituting the given body material. In the result, geometrical (vector) sum of the hoop and longitudinal stresses forms trajectories of the principal stresses in the kind of two families – left- and right-hand – cylindrical helices in the cylinder shell. Such structure of the stressed state of cylindrical specimen determines an orientation of a motion of micro deformations (including to the specimen external surface), stipulated by the vacancy-dislocation adaptation of the specimen material structure on its plastic deformation stage (the yield point zone) at the specimen tension test. In that way, a going out of the micro deformations onto the specimen external surface visualizes the trajectories of the principal stresses in the kind of Lueders lines.

The other mode of the own initial spontaneous oscillations of the finite length cylindrical specimen is the longitudinal oscillations of its parts at its ends relatively its middle cross-section. These parts are oscillated symmetrically in the opposite phases.

A variety of the previous mode of the initial spontaneous oscillations of the finite length cylindrical specimen is the torsion oscillations of its parts at its ends relatively its middle cross-section. Both these parts are also oscillated in the opposite phases.

The latter two modes determine an origin of the neck mainly in the middle cross-section of cylindrical specimen at its tension test. And the latter mode of the torsion oscillations of the specimen parts at its ends relatively its middle cross-section is also of additional interest. A feature of this mode is in that only one family of the principal stresses is loaded by the tension force during one half-period of these oscillations. The other family of the principal stresses is found to be unloaded in this half-period. During the next half-period only the other family of the principal stresses is loaded. In the other words: in the every one half-period is found to be activated only one family of spiral helices as of trajectories of the principal stresses.

The described feature determines a basic difference of the stressed state of cylindrical specimen under action of the tension force and under action of the torsion moment. In the first of these cases both the left-handed and right-handed families of trajectories of the principal stresses are simultaneously loaded. In contrast to it, in the second of these cases only one family of trajectories of the principal stresses – only the left-handed or only the right-handed – is loaded according to an action direction of the torsion moment. Comparing these cases, one can understand that the hoop tension stresses in the shell part of the tensioned cylindrical specimen and radial pressure of its shell part onto its core part will found to be twice as much than in the same specimen under its torsion under condition of equality of the principal stresses in its quantity. Accordingly, the lengthening of the specimen core part under its torsion will found to be half in comparison with its tension under the above-mentioned condition.

Side by side with the own spontaneous torsion oscillations, the cylindrical specimen can also experience the rotational oscillations of its shell part relatively its core part around its longitudinal axis without twisting. In both these cases a diameter of the specimen core part is determined by an equality of the polar moments of inertia of its core and shell parts

$$J_{pc} = J_{ps}; \quad \text{or} \quad J_p = 2J_{pc},$$

i.e.

$$p\frac{D^4}{32} = 2p\frac{D^4}{32}\left[1 - \left(\frac{d_c}{D}\right)^4\right],$$

whence

$$d_c = D/\sqrt[4]{2} \cong 0.841D,$$

where $D$, $d_c$ are diameters of the specimen and its core part correspondingly.

The numbered modes of the own spontaneous oscillations of cylindrical body are the most actual for an analysis of the stressed-strained state of cylindrical specimen at its tension and torsion tests, although such body can also experience the other modes of its own spontaneous oscillations.

The considered example of a body of cylindrical form as one from the simplest geometrical forms and the same time as a form, used very widely in the machine elements, allows, in author's opinion, to solve physically adequate the question on the stressed-strained state mechanics of the diverse form bodies under action of external forces, using the described property of self-structurization correspondingly to its form and, naturally, to its material homogeneity at its initial state.

**Discussion of results**

Effectiveness of the stated approach to an analysis of the stressed-strained state of the solid body is further demonstrated by means of the well-known results of experiments, adduced by A. Nadai in his book [5], and the some others, including own experiments of the given article author. Mechanics of the necking in the tensioned cylindrical specimen from mild steel is, in general, described in the previous article [1] of the given article author. The necessary additional explanation is in the following. From the one hand, the diagonals of the lateral sides of geometrical body in a form of a cube - the so-called external or lateral diagonals - circumscribe the linear surface in the kind of the hollowed orthogonal (right) rotary hyperboloid at its rotation around central axis of the cube. And from the other hand, the cube diagonals, passing from one of its apex to its opposite apex through its volume - the so-called internal diagonals - are inclined to the cube bases under $35^0 15' 52''$ angle. Correspondingly to it the orthogonal hyperboloid generatrices are inclined to its rotation axis and simultaneously to its bases under $45^0$ angle, and the orthogonal hyperboloid diagonals, passing through its volume, i.e. its internal diagonals, are inclined to its rotation axis under $90^0 - 35^0 15' 52'' = 54^0 44' 8''$ angle. Correspondingly, the helical trajectories of action of the principal stresses on the external surface of the tensioned cylindrical specimen are inclined to the specimen axis under $45^0$ angle, the specimen shell fracture is also realized under $45^0$ angle, and the specimen core fracture is accompanied by a tearing on the series sites, inclined mean-statistically under $\sim 55^0$ angle to the specimen longitudinal axis. The above-mentioned angles are connected between themselves by the following equality

$$\tan 35^0 15' 52'' = \sin 45^0 = \text{cosine } 45^0 = 1/\sqrt{2} = 0.707106781\ldots$$

The adduced explanation allows physically adequate to elucidate mechanics of the elastic and plastic deformation and fracture of the specimens with non-circular cross-section from mild metal at its tension test. The diversity of its cross-section forms and of a correlation of its diameters can be subdivided into the following categories:
- the specimens with a compact (solid) cross-section in the kind of a square, ellipse, rectangle with commensurable quantities of its sides or diameters;
- the flat specimens; these specimens can be, in ones turn, subdivided into two following categories:
- the flat heavy-walled specimens – with a correlation of its cross-section diameters or sides in the limits from 6 up to 10;
- the flat thin-walled specimens – with a correlation of its cross-section diameters or sides more than 10.

Morphological unity of the numbered categories of the specimens is in that any rectangular cross-section can be formed by a truncating of a cylinder at least by two parallel each other flat surfaces, placed symmetrically along the cylinder longitudinal axis. In the result, the specimens with its compact cross-section are found to be close by the cylindrical specimen, and the narrow cross-section of the thin-walled specimens is symmetrically placed relative the cylinder longitudinal axis and contains in itself the internal diagonal of the cube, inscribed into the cylinder. These geometrical features of the specimens with non-circular cross-section determine a difference of its behavior at the tension test.

Distinctive feature of plastic deformation of the specimens with a compact cross-section is in that Lueders lines, inclined under angle $45^0$ to the specimen axis, are placed only in a stripe along the mean line of the most wide side of the specimen; in the specimen with the square cross-section these stripes with Lueders lines is placed on all its lateral sides. Such localization of Lueders lines testifies to existence of cylindrical part, inscribed in the specimen volume and containing in oneself the core and shell parts. Just owing to it a fracture of such specimen starts by a forming of the neck and comes to an end by a forming of the fracture surface in a form of the "cup and cone", i.e. one is going on by the same way as one is going on at a fracture of cylindrical specimen.

During a plastic deformation of the flat thin-walled specimen under action of the tension force Lueders lines are oriented under $54^0 44'8''$ angle to the tension direction. A fracture of such specimen starts by a forming of a local thinning – the linear neck, developing from lateral edge of one of its wide sides along the straight line, coincided with one of Lueders lines and, correspondingly, with the cube internal diagonal. A fracture comes to an end when the neck reached the opposite edge of the specimen wide side.

A form of a fracture of the flat heavy-walled specimen is very sensitive to a quantity of a correlation of its cross-section sides (diameters) and one can be similar to one of the above-described two forms with symptoms of the other one. An example of such behavior of the tensioned specimen is adduced in [5].

A suitability of metal to its treatment by a pressing is determined by means of the axial compression test of cylindrical specimens with initial correlation of its height to its diameter $\bar{h} = h/d$ about 1.5 at the normal and heightened temperature. The examples of plastic deformation and the flow structure of the cylindrical specimens from mild metals, obtained by means of its longitudinal cutting, are also adduced in the book [5]. Side by side with it V.G. Osipov [6] adduced in fig. 4 the enlargement photo of a state of the cylindrical specimen, deformed plastically on 80 % of its initial height and cut across. M.E. Drits and L.N. Mogutchi [7] adduced in its fig. 4 the natural scale photo of the outward appearance of the same plastically deformed specimen. Both photos, adduced in [5, 6], show sufficiently clearly a presence of three zones:

- the central circular zone of the metal grained non-oriented microstructure by its diameter some lesser then the specimen initial diameter;
- the hoop zone along the circumference of the specimen external contour with its internal diameter ~ 0.74 of the external contour diameter of the deformed specimen; the metal grained microstructure in this zone is also non-oriented; three cracks are developed from the specimen external contour in longitudinal-radial direction with its depth nearly on all width of the hoop zone;
- the intermediate hoop zone with the clearly visible radial orientation of the specimen metal microstructure.

These three zones are also clearly visible on the butt-end flat surface of the non-cut specimen [6], i.e. a presence of the self-structuring of cylindrical specimen is also corroborated by its axial compression test.

It should be noted that the intermediate hoop zone is formed at sufficiently high velocity of the uninterrupted axial compression of the specimen not only up to its plastic deformation, but and to its flow. In the experiments, carried out by an initiative of the given article author at sufficiently low velocity of the uninterrupted axial compression, the compression force – strain diagram had a cyclical multi-stage character. Every cycle (stage) of this diagram was containing in itself two sections of the specimen deformation - elastic and plastic – in the kind of the straight inclined line and the horizontal line respectively. The end of every such cycle was a beginning of the same next cycle. In the given case, for instance, this diagram was containing in itself four such cycles. In this test the intermediate hoop zone was not formed as in every cycle as if the new specimen with its cross-section area, increased in the previous cycle, participates in the every subsequent cycle.

At the high velocity of the uninterrupted axial compression, when the deformation velocity exceeds velocity of the natural vacancy-dislocation adaptation of the specimen material stressed state, the compression force – strain diagram is realized in the kind of only one cycle. In this case the initial plastic deformation of the specimen material turns into its radial flow, forming the intermediate hoop zone between the specimen core and shell parts.

Thesis on the initial spontaneous self-structurization of the stressed-strained state of a solid body according to its form, elaborated in the author previous article [1] and developed in the given article, allows to produce mechanics of elastic and plastic deformation of cylindrical specimen from mild metal under action of the axial compression force in the following way. The specimen parts, its core and shell, having the same area of its cross-section are equally compressed by the axial pressure. At the same time a transversal – radial – deformation of the specimen core part is restricted by the hoop stiffness of the specimen shell part. Because of it the stressed state of the specimen core part is found to be close by a hydrostatic compression. At the same time the specimen shell part is under action of a combination of the hoop tension stress, aroused by a pressing onto it of the specimen core part, and of axial compression. Both these stressed states promote to the plastic deformation of the specimen from mild metal. Naturally, even if partial compensation of the hoop tension stresses in the specimen shell part by means of a loading it by lateral hydrostatic pressure allows plastically to deform even such brittle material as a marble [3]. On the other hand, a specimen from mild steel can be flattened out thus much that under action of radial pressure of the specimen core part its shell part will be tore by the few radial-axial cracks, developed from the specimen lateral contour as it is showed in fig. 4 in the mentioned article [6]. The quantitative valuation of the stressed state components of cylindrical specimen under action of axial compression corresponds, to a certain degree, to the solution method, stated in the author article [1], with the taking into account of a reverse direction of the acting forces at the specimen compression. That is the solution comes to Lame problem for the heavy-walled cylinder, but already in the three-dimensional statement. In the case of sufficiently quick axial compression of cylindrical specimen from mild steel – on a stage of its flow – the solution of Lame problem, stated in [1], must be added by the expressions allowing taking into account of the contact friction and viscosity of the specimen metal. Nadai in his book [8] has adduced the reference (as a guide) quantities of dynamical viscosity of some material, including a copper, structural steel and the others. Specialists in the field of the metal treatment by a pressing P.I. Polukhin, V.A. Tjurin, P.I. Davidkov and D.N. Vitanov in their book [9] spare the separate chapter to the contact friction of the metal half-product with the press flat plates.

The contact friction factor distinguishes very sufficiently the compression test from the tension test. The contact friction restricts, to a certain degree, radial deformation of the compressed specimen and in that way one promotes to a forming of the supporting cones - in the cylindrical specimen volume - with its bases at the specimen butt-ends. The specimen material is in a state close to a hydrostatic compression in a volume of these cones.

In his book [8] Nadai adduced the results, obtained by Meyer and Nehl (1925), of a measuring of angle $\varphi$ between longitudinal axis of cylindrical specimen and the supporting cone generatrix in dependence on the specimen relative height $\bar{h} = h/d$. In general, these results can be offered by a hyperbola, passing through points with coordinates: along the abscissa axis $\bar{h} = 0.0, 1.0, 2.5$ and along the ordinate axis $\varphi = 90^0, 55^0, 32^0$ respectively. One can see that when $\bar{h} = 1$ then the incline angle of the cone generatrix is equaled to the angle between the cube rotary axis and its internal diagonal as the action trajectory of the principal stresses in the specimen volume. Axial compression of the short specimens – with $\bar{h} < 1$ – leads to a superposition of apexes of the supporting cones and, correspondingly, to an increase of the specimen resistance to the compression force, applied to the specimen. Axial compression of the long specimen – with $\bar{h} > 1$ – leads to a transformation of these cones into the cylindrical core part of the specimen.

S.P. Timoshenko [3] adduced the results of experimental research, carried out by C. Sachs (1905), in the kind of the specimen relative height ($\bar{h}$) – compressive stress diagram. An increase of $\bar{h}$ in the limits from 0.27 up to 2.0 is accompanied by a decrease approximately in two times of the compression stress. These cylindrical specimens were made from mild copper of the same diameter. Axial compression of the cylindrical specimens by the preliminary lubricated plates of the test machine [3] allows to except practically a formation of the above-described supporting conical zones with the stressed state close by hydrostatical compression. In this case the cylindrical core both in the cylindrical and cubic specimen arouses the hoop tension stresses under action of its axial compression. In this situation a character of a deforming and of a fracturing of the specimen is determined by its material properties, i.e. by its plasticity or brittleness. The brittleness degree is usually determined by a ratio of the compression strength to the tension strength of the given material. Under action of axial compression the specimen from mild metal is flattened, the specimen from brittle metal (grey iron) is fractured along the diagonal surface, formed by two cracks, developed along the both opposite trajectories of the principal stresses on its lateral surface, and the specimen from stone (granite, cement) is fractured along the longitudinal cracks.
Considering the stressed state of a solid body under action of axial compression, one cannot but mention an influence of the body form on its load-carrying ability. Nadai [5], on a base of the experimental research results on axial compression of several modifications of the piped specimens from the burnt porcelain, carried out by Westinghouse Laboratories (May,1941), offered a form of the piped specimen with its mean surface in the kind of the hollowed hyperboloid of rotation with its thickness, increased proportionally to its diameters. A question arises in connection with it: what one from multitude of the hollowed hyperboloids of rotation meets the requirement of the most strength in a combination with the least weight? In his previous article [1] the author showed that the trajectories of the principal stresses in the shell part of cylindrical specimen form two orthogonal families – left- and right-handed – of cylindrical helices, and on reaching by the specimen of its ultimate strength these helices are straightening oneself and form the neck in the kind of the hollowed orthogonal (right) hyperboloid of rotation. The incline angle of the straightened generatrices of such hyperboloid to its rotation axis is equaled to the incline angle of the helical trajectories of the principal stresses on the specimen external surface and one makes up $45^0$. At the same time, the examples of a test of the specimens from brittle materials on axial compression, adduced in the book [5], showed that Lueders lines, as reflection of trajectories of the principal stresses, are inclined on the greater then $45^0$ angle. In fig. 301 (photo), just mention book, the incline angle of Lueders lines on external surface of the cylindrical specimen from marble makes up $60^0$. If one take that a ratio of the compression strength to the tension strength is equal for marble and the burnt porcelain, then it should be took $90^0 - 60^0 = 30^0$ as the incline angle of the straight generatrices of the hyperboloid mean surface to its rotation axis. The greater thickness of such hyperboloid body is to have been in its least cross-section (neck) and one has to meet the equality condition of the wall cross-section area to the orifice cross-section area in the hyperboloid neck. This quantity of the wall cross-section area should be taken for the rest cross-sections of the hyperboloid, therefore its wall thickness should be decreased proportionally to an increasing of its diameter. As to the hoop stresses, mentioned by Nadai, the hollowed hyperboloid of rotation not contains such component of its stressed state on a definition of this figure. The author explained this feature in his previous article [1].

Completing a consideration of the solid body state under action of axial compression, the author supposes to be expedient a considering of the question, connected with the high-speed and super-high-speed action of compressive forces on the elastic-plastic body, capable to a flow.

The simple example of the quite high-speed action of compressive force onto a body can be presented in the kind of the blows with usual bench hammer onto a flat butt-end of cylindrical bar from mild metal (aluminum, copper, steel). Under such action, when the effective contact duration of the hammer with the bar is smaller than duration of a running of the compression longitudinal sonic wave in the shortest from these two bodies in the direct and reverse directions, the sufficient part of the hammer kinetic energy is transformed into the local plastic deformation of the bar. Action of such many blows leads to a spilling of the bar butt-end up to an origin of radial cracks, developed from external contour of this part of the bar. A thickness of the plastic deformation zone in the bar – along its axis – can be determined by expression

$$\boldsymbol{d}_{pl} = l_h \cdot t_i / t_{swh},$$

where $l_h$ is a length of the shortest body, participated in the blow interaction, $t_i$ is duration of the blow impulse, $t_{swh}$ is duration of a running of the direct and reverse longitudinal sonic wave in the shortest body; in the given case, for instant, the hammer is accepted as the shortest body.

A thickness of the local plastic deformation zone not exceed usually a half the bar diameter, and a ratio of the axial compression stresses on the ends of this zone corresponds to a ratio of acoustic stiffness of the bodies, participating in the blow interaction

$$\pmb{s}_0/\pmb{s}_d = \pmb{r}_h \cdot c_h \cdot A_h / (\pmb{r}_b \cdot c_b \cdot A_b),$$

Where $\pmb{s}_0$ and $\pmb{s}_d$ are axial compressive stress at the bar butt-end and on a distance $\pmb{d}_{pl}$ from it, respectively, $\pmb{r}_h, c_h, A_h$ and $\pmb{r}_b, c_b, A_b$ are density, the sonic wave velocity and the cross-section area of the hammer and the bar, respectively. A product $\pmb{r} \cdot c$ is accepted to be called by the acoustic stiffness of the given material or its impedance, and a product $\pmb{r} \cdot c \cdot A$ is offered [10] to be called by the body blow stiffness. If in the considered case one will suppose that the hammer and bar impedances are close by or equal each other then previous equality takes the kind

$$\pmb{s}_0/\pmb{s}_d = A_h/A_b.$$

Let us suppose that the hammer cross-section area is sufficiently larger then that of the bar. Than the axial compressive stress at the bar butt-end will be larger then that on a distance $\pmb{d}$ from its butt-end. Therefore the spilling of the bar under the hammer blows is going on just at the bar butt-end. Under the hammer every blow the bar core part, experiencing the longitudinal compressive impulse, exerts the impulsive radial pressure onto the bar shell part. Under action of the core radial pressure in combination with own axial impulsive pressure the bar shell part is spilt and one forms the circular collar, increasing its diameter with the hammer every blow. By a sufficiently quantity of the hammer blows, the radial spilling of the bar collar is accompanied by origin in it of radial cracks, localized more or less uniformly along the collar circular contour.

The adduced example of the hammer and bar interaction can be compared with the following example. A water drop falls freely from height up to one meter onto the flat solid surface. Supposing that the water drop as any spherical body contains in itself the core and shell parts with the core radius ~ 79.4 % and a thickness of its shell part ~ 20.6 % of the drop external radius, as it is showed in the given article Solution section, the transient process of the drop interaction with the solid obstacle can be presented in the following sequence. In a beginning of the drop contact with the obstacle, the drop spherical shell is deformed against the flat obstacle. At the same time the spherical core, because of its braking, squashes under itself the drop shell part, arousing a transversal radial spreading of this part of the drop shell. The drop core contact with the flat obstacle is also accompanied by its flattening and by its transversal radial spreading. The drop shell part, placed over the core part and round it, flows round the drop deformed core and begins its transversal radial spreading under action of its own braking and of a transversal radial pressing of the drop core onto it. The described period is the period of transformation of the longitudinal momentum into its transversal radial momentum, accompanied by a formation of the circular water wave on the flat solid surface. In the next period this water wave under action of its hoop compression begins its radial motion – from away the drop fall centre and to the drop fall centre. The wave part, moving to the centre, forms the thin vertical fountain. The wave other part spreads, forming the compact circular front contour. Such motion is accompanied by the resistance forces in the kind of the water internal friction and the water surface tension. The hoop compression of the spreading wave is decreased as its radius is increased, and once this compression is decreased up to naught and further one transforms into the hoop tension. Then the compact circular front contour of the spreading wave disintegrates to the series of the same separate jets. Diameter of these jets is decreased gradually under action of the water surface tension, and its stop is accompanied by elastic recoil with a throwing out the micro drops out of an end of every jets. H.E. Edgerton (1934) had made the well-known wonderful photo of the disintegration final stage of the milk falling drop onto a plate.

Comparison of the described processes of the blow action on the bar from metal and on the water drop testifies to that a motion of mechanical energy in a solid body and fluid and also its accumulation and distribution in these bodies and fluids, correspondingly its form, and transmission of mechanical energy in the contact interaction of the bodies and fluids between themselves – all these processes have the same dynamical – wave – nature. Difference in a behavior of such material as metal from fluid under action of external forces is stipulated by a presence of static Young's modulus and the very high viscosity in the solid body in contrast to such fluids as air, water. This thesis, to a certain degree, is corroborated by H. Kolsly and D. Rader experiments [11] with the bars from polymethilmethacrilat (PMMA) and polystyrene and also with water, put to an action of impulsive load. M.A. Lavrentjev and B.V. Shabat [12] adduced photos of longitudinal section of the welded joints all over its contact surface of two parts in the kind of a conic bar and adjacent to it a shell and also of two sheets from mild metal, are made by means of the explosive, covering the external surface of one of the joined parts. The vortex-wave structure of these welded joints corresponds exactly to the vortex-wave structure, arising in relative motion of two water (or air) stream, contacting each other along initial flat surface. The evolution scheme of such vortex-wave structure development is presented in fig. 7 in author's article [13]. The same structure, accompanying the circular interaction of the air streams, is adduced by M.D. Van Dyke in fig. 81 (photo) in his An Album [14]. The super-high-speed action of the shaped charge explosion compresses its conical metallic shell and transforms it into the high-speed jet from the unmelted metal [12]. Such metallic jet goes through metallic barrier by means of its washing away without its melting. Now it is well-known the technology of a cutting by the liquid (water) jet under high pressure (up to 5,000 atmospheres) of the different solid bodies, including metals.

The physical adequacy will not be broke by comparison of the water cavitation in the converged-diverged nozzle (Venturi nozzle) with fracture of the core part of cylindrical specimen from mild steel under its axial tension, in author's article [1]. In both these cases material – water, steel – is in its state close by hydrostatic tension.

Let us compare the water stream state in the pipe with the stressed state of cylindrical bar from steel under action of axial force. The water steady motion in the pipe is going on under action onto it of the invariable force of a pressure drop and one is determined by the water internal friction – viscosity – against the pipe wall. The internal friction as a dynamical factor determines, in one's turn, the static pressure distribution along and across the stream, containing the axial and two transversal – radial and hoop – static components of the stream total head vector (STHV) [15]. Geometrical sum of these static components with the dynamical component in the kind of the stream velocity head determines completely the stream energy state in any cross-section of the stream. In this situation one cannot but pay attention to a dynamical character of the stream in the kind of its motion and in the same time to the pressure static character along and across of the same stream. I.e. the stream moves, but pressure in it "is stopping". In contrast to it, cylindrical bar (specimen) and, correspondingly, its stressed state remain immovable at its testing on axial tension or compression, at least on stage of its elastic deformation. Such difference in behavior of the water stream in the pipe and metallic specimen is connected with a lack of Young's modulus in water, owing to that water possesses always additional degree of freedom in the kind of its infrastructural mobility. Therefore the water stream in pipe can be moved and in the same time its stressed state can be immovable. This static stressed state in the water stream is quite equivalently to the stressed state of metallic bar under action of longitudinal force, stipulated by its own weight. The components of the total head vector of the water stream in pipe form in this stream, under action of its friction against the pipe wall, the same two families of trajectories of the static pressure action in the kind of cylindrical helices as the trajectories of the principal stresses in the steel cylindrical bar under its axial compression by its own weight forces.

A fitness of the machine elements to transmission of the torsion moment is determined by the test results of mainly cylindrical specimens by means of the torsion-testing machines. In his book [5] Nadai dedicated to this kind of a testing the separate chapter, in which he adduced two methods of a calculating of the shear (tangential) stresses in cylindrical specimen under action of the torsion moment. Side by side with it Nadai adduced some results of experiments, carried out, in particular.. by J.H. Pointing (1909, 1912), E.A. Davis (1937) and H.W. Swift (1947). These results, in contrast to the theoretical finding, testify to the fact that the specimens from the piano wire and in the kind of cylindrical bar from mild copper, brass and aluminum lengthen oneself under action of the torsion moment. Swift showed also that a change of the moment direction is accompanied, in the beginning, by its shortening according to a decreasing of the torsion moment of initial direction and than by the new lengthening according to an increasing of the reverse direction moment. Davis detected experimentally an origin of the cold-work strengthening in the core of cylindrical specimen from mild copper in the result of its plastic twist. He ascertained that the hardness rises as an approaching to the specimen cross-section centre. On these grounds Nadai surmised a possibility of existence of normal (principal) – axial, radial and hoop – stresses in cylindrical specimen under action of the torsion moment, but he did not develop this idea. This idea is not developed till now neither in the theory of elasticity nor in the theory of plasticity, although solution of a problem on the shear stresses in this kind of a loading can be dated, according to Nadai [5], by 1882.

To elucidate a question on possibility of existence and of a distribution character of the principal stresses in cylindrical specimen from statistically isotropic material under action of the torsion moment, it should be took into account the well-known equivalence of the stressed state in a pure shear and in combination of normal (principal) stresses. This equivalence allows presenting the stressed state of cylindrical specimen under action of torsion moment both in the kind of a pure shear and in the kind of combination of the principal stresses as substituting mutually each other. An application in specific case of either of approach to the specimen stressed state analysis is determined by the specimen material properties. In particular, a taking into account of normal stresses, determining the kind of the cylindrical specimen fracture under action of the torsion moment, is to be necessary for the brittle material, having its tension strength in 2…5 times less than its compression strength. In this case it is necessary to consider the cylindrical specimen stressed state by means of the principal – tensioning – stresses, acting along the cylindrical helices, having its angle equal $45^0$ and its direction according to the direction of the torsion moment action. When the tension stress along this one family of the helices reaches the tension strength of the brittle material, then such specimen is fracturing on the helical surface with its external helical edge, oriented orthogonally to the tension stress trajectory. Although the specimen shell part, underwent to an action of the hoop tension stress, exert radial pressure onto the specimen core part, a lengthening of such specimen is insignificant because of the low tension strength of brittle material.

To elucidate the stress state character and the fracture shape of cylindrical specimen from sufficiently plastic material under action of the torsion moment, the given article author has carried out several experiments in the kind of a twisting by hand of the wire pieces from mild aluminum (of type 2L4 British Standards Specifications) with its diameter 1.8 mm. These experiments showed that a fracture of such specimens was going on after that as the incline angle of helical lines (Lueders lines) reached ~ $20^0$ on the wire surface. I.e., in the given case the hoop tension stresses in the specimen shell part and, correspondingly, its radial pressure onto the specimen core part are found to be in ~ 2.5 times greater than under $45^0$ angle. In that way the specimen shell part by its thickness ~ 16 % of the specimen radius is under action of the hoop and axial tension stresses in its correlation, determined by a tangent of ~ $20^0$ angle of the helical trajectories of the principal stresses and also of internal pressure, stipulated by the core part stiffness. In one's turn, the specimen core part is in lateral compression state under action of onto it of the shell part and also in the shear state under action of its twisting. In this situation one half the torsion moment energy, applied to the specimen, is realized in the kind of the shell part stressed state and the energy other half is realized in the kind of the core part stressed state in accordance with equality of the polar moments of inertia of the specimen core and shell parts.

The tested specimen fracture begins always by the partial rupture of one of the helical coils in the specimen shell part and then one is accompanied by the specimen transversal cutting by means of mutual turn of the specimen butt-ends, adjoined each other in the cut flatness, around the rest intact part of the partially ruptured helical coil. In that way the specimen shell part with its relative thickness ~ 16 % and its cross-section relative area ~ 29.3 % is stretched as a cylindrical spring under action of the torsion moment and one squeezes the specimen core part with its cross-section area ~ 70.7 %. In the result the specimen core part is lengthened – in the beginning elastically, then plastically – and one carries along together with oneself the specimen shell part, arising thereby the specimen lengthening. In the case of the specimen from mild copper these conditions will arise the cold-work hardening of the specimen core part, detected by Davis in his above-mentioned experiment. The described lengthening of the specimen promotes to an increasing of the helix angle in its shell part. The above-mentioned initial partial rupture of one of the helix coils is going on just in a moment, when a decreasing of the helix angle under action of the torsion moment is balanced by an increasing of this angle because of the specimen lengthening. Strong influence of the specimen lengthening onto an increasing of the helix angle and, correspondingly, of the shell axial tension stress is stipulated by that the core part cross-section area is more in 70.7 / 29.3 = 2.4… times than the shell part cross-section area.

In the above-described experiments of the given article author the aluminum wire fracture under action of the torsion moment was not accompanied by the visible residual change of the specimen form or by its visible residual local deformations as it takes place at the cylindrical specimen tension or compression test. The only shear character of such fracture and small quantity of the helix coil angle - $20^0$ - testify to a high ductility of the commercial pure aluminum. In this connection the author supposes advisable to recommend to a plasticity valuation of any material by means of the helix angle (Lueders line angle) at the cylindrical specimen test by the torsion moment. The approximate correspondences are the followings:

$j \leq 20^0$ - high-ductile material (the commercial pure aluminum);
$20^0 < j \leq 30^0$ - ductile material;
$30^0 \leq j \leq 40^0$ - low-ductile material;
$j = 45^0$ - brittle material;
$45^0 \leq j \leq 75^0$ - high-brittle and/or anisotropic material (timber).

The ascertain of existence and the elucidation of the principal – normal – stresses and its orthogonal components in cylindrical specimen under action of the torsion moment allows also to discover nature of the stress concentration under action of the torsion moment in zone of the specimen diameter sharp change. A quantity of the principal stresses and, correspondingly, of its components in the specimen part of the lesser diameter is found to be larger then that is in the specimen part of the grater diameter proportionally to a correlation of its polar (torsion) moments of resistance, i.e. to $(D_1/D_2)^3$. In the other words, even at the not great difference in diameter of these parts of the specimen the quantities of the principal stresses and, correspondingly, its components are found, in the specimen lesser diameter part, to be so larger that the stressed state of the specimen greater diameter part can be left out of account. Therefore below it is considered the stressed state only in the specimen part of the lesser diameter, contained, in one's turn, its core and shell parts. The transversal component of the principal stresses in the kind of the hoop tension stress, acting in the shell part, squeezes the core part and provokes its lengthening. In the result the shell part is loaded by axial tension stress in addition to the axial tension stress, as a component of the principal stresses. Because of its lengthening, the core part exerts axial pressure onto the specimen greater diameter part and thereby one provokes the local additional tension stress in the shell part. Although the axial compression force in the core part is balanced by the axial tension force in the shell part, the local axial tension stress in the latter is more in 2.4 times than the local axial compression stressing the core part. In that way the shell part is loaded by a sum of the following three components of the axial tension stress: as the principal stress component, as the component, stipulated by the core part lengthening, and as the component, stipulated by a pressing of the core part onto the specimen greater diameter part. The first two from these components are distributed uniformly along the specimen axis; the third component has the local character. Ratio of a sum of these three components to a sum of the first two components is the so-called stress concentration factor, stipulated by the torsion moment action, applied to cylindrical specimen, having a sharp change of its diameter. In that way the stress concentration is retribution for non-observance of the smooth conjugation of trajectories of the principal stresses.

**Final remarks**
The approach and the problem solution, stated in the given article, and the examples, adduced in it, are practical realization of Aristotle thesis on necessity to develop our knowledge from the more intelligible and evident for us to the more intelligible for nature as the intelligible for us and the intelligible in general are not the same ones. This Aristotle thesis is highly urgent the present day as the modern mechanics is highly far from being perfect.

**Acknowledgements**
The author wants to express the deep gratitude to the governing body and officials of International Biographical Centre, Great Britain, to the governing body and officials of American Biographical Institute, Inc., U.S.A. and also to librarians of scientific and technical libraries, rendering to the author the high encouragement and passing the modern knowledge to next generation of specialists._________________________________________________

[1] S.L. Arsenjev, "To the strength first problem full solution: mechanics of a necking", http: //uk.arXiv.org/abs/physics/0609149, 2006
[2] F.A. McClintock, Ali S. Argon, Mechanical Behavior of Materials, Addison-Wesley Publishing Company, Inc., Reading, Massachusetts, U.S.A. and Addison-Wesley (Canada) Limited, Don Mills, Ontario, 1966 – Transl. from Engl. into Rus., Mir Publishing, Moscow, 1970
[3] S.P. Timoshenko, Strength of Materials, part II, $3^{rd}$ Ed., D. Van Nostrand Company, Inc., Toronto – London – New York, 1956, - Transl. from Engl. into Rus., Nauka Publishing, Moscow, 1965
[4] F.S. Tse, I.E. Morse and R.T. Hinkle, Mechanical Vibrations, Allyn and Bacon, Inc., Boston, 1963 – Transl. from Engl. into Rus., Mashinostroenie Publishing, Moscow, 1966
[5] A. Nadai, Theory of Flow and Fracture of Solids, $2^{nd}$ Ed., McGrow – Hill Book Company, Inc., New-York, 1950 - Transl. from Engl. into Rus., Vol. I, The State Publishing of Foreign Literature, Moscow, 1954
[6] V.G. Osipov, "Process of fracture by shear at simple compression and tension", In Collected Articles on the Strength of Materials, USSR Academy of Sciences Publishing, Moscow, 1956
[7] M.E. Drits, L.N. Mogutchi, "Mechanical properties of the deformable alloy MA9 at high temperature", Ibid
[8] A. Nadai, Theory of Flow and Fracture of Solids", Vol. 2, McGrow – Hill Book Company,Inc., New York – Toronto – London, 1963 - Transl. from Engl. into Rus., Mir Publishing, Moscow, 1969
[9] P.I.Polukhin and all, Treatment of Metals by Pressing in Machinebuilding, Mashinostroenie Publishing, Moscow, 1983
[10] E.V. Alexandrov,V.B. Sokolinski, The Applied Theory and Calculations of Blow Systems, Nauka Publishing, Moscow, 1969
[11] H. Kolsky, D. Rader, "Stress waves and fracture", in Fracture, Vol. I , Ed. By Liebowitz, Academic Press, New York – London, 1968 - Transl. from Engl. into Rus., Mir Publishing, Moscow, 1973
[12] M.A. Lavrentjev, B.V. Shabat, The Problems of Hydrodynamics and its Mathematical Models, Nauka Publishing, Moscow, 1977
[13] S.L. Arsenjev, "To the fluid motion dynamics", http: //uk.arXiv.org/abs/physics/0605175, 2006
[14] M.D. Van Dyke, An Album of Fluid Motion, The Parabolic Press, Stanford, CA, - Transl. from Engl. into Rus., Mir Publishing, Moscow, 1986
[15] S.L. Arsenjev, "Fluid motion physics: the total head vector of the real fluid stream", http: //uk.arXiv.org/abs/physics.flu-dyn/0711.1705, 2007